\documentclass[12pt,preprint]{aastex}
\usepackage{graphicx}
\begin{document}

\title{What do exotic equations of state have to offer?%\thanks{Grants or other notes
}


\author{J.E. Horvath}


\affil{J.E. Horvath \\
              IAG, Department of Astronomy \\
              Universidade de S\~ao Paulo, \\
              Rua do Mat\~ao 1226, 05508-900 S\~ao Paulo SP, Brazil\\
              Tel.: +55-11-3091 2806\\
              \email{foton@astro.iag.usp.br}           %  \\
}

\begin{abstract}
We present a short general overview of the main features of exotic
models of neutron stars, focusing on the structural and dynamical
predictions derived from them. In particular, we discuss the
presence of ``normal'' quark matter and Color-Flavor Locked (CFL)
states, including their possible self-bound versions, and mention
some different proposals emerging from the study of QCD
microphysics. A connection with actual observed data is the main
goal to be addressed at this talk and along the meeting. It is
demonstrated that exotic equations of state are {\it not} soft if
the vacuum contributions are large enough, and argued that recent
measurements of high pulsar masses ($M \geq \, 2 \, M_{\odot}$)
create problems for {\it hadronic} models in which hyperons should
be present.
\end{abstract}

\section{Introduction}

The quest of the internal composition of compact stars has been
going on for decades, in close connection with the work in nuclear
and particle physics. Although deemed in some sense ``simpler"
than magnetospheric phenomena, dealing with matter at the extreme
conditions inside pulsars/neutron stars has never been easy. As
2006, the consensus about the nature of matter at several times
the nuclear saturation density is weak, if anything. Several
phases/components of the nuclear fluid have been proposed and
studied, but decisive evidence for or against them is hard to
obtain. In fact, and as discussed several times during this
meeting, advances on the observational side has allowed one for
the first time to probe key macroscopic properties of compact
stars (masses, radii and a few others) that reflect the internal
composition indirectly. However, a few remarks on this last
statement are in order: on the one hand the precision attained by
measurements has improved greatly but {\it not} to the point far
beyond any suspicion, and on the other hand, supranuclear
components with poorly constrained parameters (coupling constants,
vacuum expectation values, etc.) will not be ruled out, even by
very precise observations. Therefore, work is needed on the
theoretical and terrestrial laboratories as well.

Overall, I believe it is fair to state that the bulk structural
properties are well-known in the subnuclear domain (more strictly,
below the neutron drip density $\sim \, 10^{11} \, g \, cm^{-3}$).
However, important questions involving magnetic fields still
remain (van Adelsberg et al. 2005). This is quite important to
settle since it is where the star surface is seen by experiments
due to electromagnetic radiation (see discussions about spectral
lines and related topics in this meeting). There is also important
information in timing irregularities, most notably glitches,
believed to originate at the inner crust (although the
``conventional wisdom'' has been recently challenged, see the
contribution by B. Link in this meeting and references therein)
However, deep below the stellar crust it is increasingly difficult
to construct a clear picture of the composition and therefore of
the stellar structure. Condensates ($\pi^{-}$, $K^{-}$, etc) are
still possible depending on microphysics and drastically alter the
thermal and dynamical behavior of the star. Fundamental degrees of
freedom (i.e. quarks) may also constitute the main part of the
core, and even most of the star if they happen to be of the
``self-bound'' type (i.e. do not decay back to ordinary nuclear
species once formed). Recently, a lot of attention has been paid
to paired quark matter in a variety of phases still being studied.
A recent summary of these matters can be found in Weber (2005).
Hereafter we shall mainly concentrate on the self-bound phases
because they bring the greatest modifications to the structure and
they remain viable alternatives to the supranuclear matter
composition.

\section{Strange quark matter and its paired version}

Strange quark matter (SQM) is an extreme version of a cold quark
plasma in which, by hypothesis, the energy per baryon number unit
is selected (with the chosen parameters) to fall below the mass of
the nucleon. This possibility of having matter so strongly bound
that it does not wish to return to the normal hadronic state was
first discussed by Bodmer (1971), rediscovered by Terazawa (1979)
and finally relaunched colorfully by Witten (1984) more than 20
years ago. Many studies, both experimental and theoretical,
devoted to the SQM hypothesis have produced interesting results
and some controversial arguments {\it against} its existence, but
with loopholes in them. A few candidates have appeared to SQM in
cosmic rays (Bj\"orken \& McLerran 1979, Ichimura, Kamioka,
Kitazawa et al. 1993, Choutko 2003, Madsen 2005) with fluxes
consistent with astrophysical injection scenarios (i.e. merging of
compact stars, supernovae). SQM formation on $\tau \leq 1 \, s$
timescale has been studied (Benvenuto \& Horvath 1989, Lugones,
Benvenuto \& Vucetich 1994, Dai, Peng \& Lu 1995) and tentatively
related to core-collapse supernovae, perhaps driven by photons
(Chen \& Xu 2006) instead of mechanical energy transfer or
neutrinos.

The existence of SQM would be important for compact objects, since
within this picture {\it all} of them should be ``strange stars''
instead of neutron stars. However, there is still the issue of the
timescale for the conversion, since while in supernovae models the
latter is quite short, it could be stretched by several orders of
magnitude depending on microphysical details (see, for example,
Lugones \& Bombaci 2005). For the static, non-rotating structure,
strange stars are constructed by integrating the
Tolman-Oppenheimer-Volkoff equation with an equation of state of
the form

\begin{equation}
P={\frac{1}{3}}(\rho - 4B)
\end{equation}

which has been extensively used because of its proximity with more
detailed calculations including the finite s-quark mass and
quark-quark interactions. The importance of the vacuum term, here
written as $4B$ in the spirit of the well-known MIT bag model
(Degrand, Jaffe, Johnson \& Kiskis 1975) which produces a zero
point pressure at finite (and large) energy density can not be
overstated: the Bodmer-Witten-Terazawa hypothesis would preclude a
``normal'' matter crust in contact with it (thus limiting its
total mass) unless a structured form of the quark matter itself is
present (Benvenuto, Horvath \& Vucetich 1990, Heiselberg, Pethick
\& Staubo 1993, Alford, Rajagopal, Reddy \& Steiner 2006) and may
be responsible for phenomena commonly attributed to the inner
crust (i.e. glitches). The alternative is a ``floating'' normal
crust supported by electrostatic forces, and then necessarily
quite light, perhaps too light to produce the observed
phenomenology.

Quite independently of these considerations, there is a widespread
belief that, because of its underlying free quark derivation, an
equation of state (EOS) like eq.(1) must be very soft. This is far
from being true: the issue of the softness/stiffness is rather
related to the vacuum energy term, the real agent which determines
the hardness of the EOS. We shall give examples of this behavior
when discussing the stellar models.

A lot of activity has recently been seen on the effects of pairing
interactions in dense quark matter. The issue is not new, since in
the early '80s a few works addressed the
superfluid/superconducting properties of paired quarks (Bailin \&
Love 1984 and references therein). However, those approaches were
based on perturbative schemes, and therefore obtained (quite
consistently) gaps of the order of $1 \, MeV$ or so, much smaller
than the natural scales of the problem (say, the quark chemical
potential). The recent works (Alford \& Cowan 2006, G\'omez Dumm,
Blaschke, Grunfeld \& Scoccola 2006 and references therein) have
tried to calculate the phase diagram more directly, without
resorting to perturbative schemes. As a result, several pairing
possibilities ($u$ and $d$ quarks only, 2SC phase; all $u,d$ and
$s$ quarks at a common Fermi momentum -not energy!-, the CFL
state) were found with gaps as large as $100 \, MeV$. Other
possibilities, like a gapless CFL phase or the solid-like
Larkin-Ovchinnikov-Fulde-Ferrell (LOFF) state are being considered
and reflect a high complexity of the QCD phase diagram that might
be important for compact stars (Ruster, Shovkovy \& Rischke 2004).

While this task continues, it is perhaps worthwhile to remark the
importance of pairing energies for the stability issue discussed
above: it is found in simple models (Lugones \& Horvath 2003) that
the stability window (i.e. the place in parameter space inside
which paired CFL matter would be stable) is greatly enhanced when
compared with the same parameters for unpaired matter. Fig. 1
displays the situation for a model with constant gaps which were
varied within the expected range. Perhaps pairing is a big clue to
the ground state of matter relevant to astrophysics after all.
This is the meaning of the (unimaginative) name ``CFL strange
matter''.

\section{Effects on stellar models}

The general trend of stellar models calculated with self-bound
equations of state is well-known: in sharp contrast with neutron
matter calculations, for which $R$ grows to $\sim 100 \, km$ for
finite small baryonic mass values, self-bound stars can be found,
in principle, down to tennis-ball sizes continuously (i.e. $R
\rightarrow 0$ when $M \rightarrow 0$) and beyond, inside the
realm of ``strangelets''. This is because binding comes from
strong interactions and not from gravity. Of course, there {\it
is} a Chandrasekhar mass for SQM or CFL strange matter sequences,
which brings us to the question of the vacuum energy again.

In its simplest form of eq.(1) it is well-known that stellar
models at the maximum mass scale as $B^{1/2}$. This property is
related to the linearity of the EOS, and holds approximately if
the latter is not strict. When pairing energy is present, the free
energy of the paired mixture $\Omega_{CFL}$ is smaller than the
unpaired version by a term quadratic both in the gap $\Delta$ and
the chemical potential $\mu$ (Alford \& Reddy 2003)

\begin{equation}
\Omega_{CFL} = \Omega_{free} - \frac{3}{\pi^2} \Delta^2 \mu^2 + B
\end{equation}

All the important thermodynamic quantities can be derived from
eq.(2), which provides an equation of state for the CFL mixture.
With that ingredient it is immediate to calculate stellar
sequences of cold stars composed by this self-bound version of
quarks. In this approach the gap has been assumed as a constant,
in fact theoretical expectations strongly suggest that a
functional dependence ensures, but at this time it is not possible
to state anything reasonable and quantitative about its nature.
The maximum mass along the sequence is shown in Fig. 2, and
increases with increasing gaps. Of course, there must be an upper
limit to the pairing energy gain in nature, otherwise all matter
would decay into the more bound state (as pointed out by P.
Haensel during the meeting), but this limit is not obvious and
must be calculated consistently for each considered model (see
Fig. 1). One important point to note here is the rather high
values for the maximum mass along the sequence that can be
obtained for gap values deemed quite modest (i.e. $M_{max} \geq 2
\, M_{\odot}$ for $\Delta \sim \, 100 \, MeV$). This means that
the EOS is {\it not} soft, but rather stiff whenever the effective
vacuum energy is large enough (composed in these models by a
combination of the true vacuum and the condensation energy
together).

With the use of analytical general relativistic solutions (Delgaty
\& Lake 1998) one can go further and find the {\it locus} of mass
maxima as a function of the radius $R_{max}$. The answer is a
curve indicating that, in general, larger maximum masses of
self-bound sequences must have increasingly larger radii.
Therefore, the observations of large compact star masses can
potentially set a lower limit to their radii as well. This test is
perhaps one of the simplest to perform since masses can be in some
cases obtained with great precision, whereas radii are somewhat
more indirect.

\section{Masses and radii: recent observations}

The zoo of compact star masses and radii has been growing
recently, and there is now a firm expectation of finding more
reliable limits to the internal structure than hitherto possible.
From this point of view, important determinations are those of $M
\gg 1.4 \, M_{\odot}$ and $M \ll 1.4 \, M_{\odot}$, because this
are the limits where the self-bound and conventional models are
more different. Nice, Splaver, Stairs et al.(2005) claim of $M =
2.1 \pm 0.2 \, M_{\odot}$ ($M =2.1 ^{0.4}_{0.5} \, M_{\odot}$ at
$2-\sigma$ level) for the compact star designated as PSR
J0751+1807, and the Baker, Norton \& Quaintrell (2005)
determination of $M = 0.91 \pm 0.08 \, M_{\odot}$ for SMC X-1 are
just two examples of this ``spreading'' around the older canonical
value of $1.4 \, M_{\odot}$.

In fact this is one of the main reasons of why should we care
about self-bound models: while it is generally believed that high
masses disfavor a quark composition, it could be that {\it
hadronic} models have a serious problem with them. This is because
the appearance of hyperons (known to exist for decades) generally
soften the equation of state below $1.4 \, M_{\odot}$ or so, a
result found consistently over the years from microphysical
approaches. The existence of quark cores does not help either
because the maximum mass decreases with respect to the purely
hadronic model. Therefore, either hyperons couple to neutrons and
protons with strengths capable of giving a large extra repulsion
(thus rising the maximum mass of the sequence), {\it or} it is
acknowledged that truly exotic models, like the self-bound ones,
are more compatible with the high masses (the problem with
hyperons has been noted before by M. Baldo, F. Burgio and
coworkers). But this extreme possibility would also predict that
the big difference in radii between conventional and self-bound
models would begin below $0.5 \, M_{\odot}$, not around $1 \,
M_{\odot}$, and it is unclear how and if such low-mass stars are
formed in nature. This finally means that $\sim \, 1 \, M_{\odot}$
stars should show radii around $10 \, km$, not $6-7 \, km$ as
previously thought. While we can not prove that exotic matter is
present in compact stars, it is also not guaranteed that
microscopic EOS with all the degrees of freedom known from
laboratories can fit the observations either.

\section{Conclusions}

We end this brief exposition about some features of self-bound
models by saying that understanding of the {\it vacuum} is the
real clue for advances in dense matter physics. This is not unlike
other fields of physics, like the well-known crisis in cosmology
about what the quantum vacuum should be and what actually is
(Freedman \& Turner 2003). The same vacuum issue, but related to
matter well above the saturation density, looks even more
formidable, and its understanding should solve in the wash the
issues of the existence of self-bound states and the features of
the resulting EOS. It is also important to remark again that there
is no hadronic model devoid of problems with the high mass end:
either they ignore hyperons or are solved in a mean field approach
or some other scheme with its own problems and questions. There
is, however, the possibility of an extreme stiffness of the
equation of state, such as the hyperons do not appear at all,
because the relatively low density at the center. The good news is
that we can now foresee actual tests of the stiffness of the EOS
in the near future.

As stated, this stiffness is very important to establish, because
the very introduction of $\Lambda$ particles and other hyperons
would then call for very exotic interactions among them at least.
Otherwise the resulting EOS become so soft that measured masses
around $2 M_{\odot}$ happen to lie well above the maximum masses
of the respective theoretical sequences. This is why strongly
repulsive interactions would be required if hyperons appear inside
compact stars. High masses may be {\it pointing} towards the
exotica rather than excluding them (\"Ozel 2006, Alford, Blaschke,
Drago et al. 2006).

To conclude, we would like to quote an inspiring sentence from the
English literature that may (or may not) be related to these
topics, in which a bit of fantasy is always hidden
\medskip

{\it Horatio, there are more things in Heaven and Earth}

{\it Than are dreamt of in your philosophy}

\hfill{Hamlet, Act I, Scene V}

\section{Acknowledgements} We would like to acknowledge the
financial support of CNPq (Brazil) and the Organizing Committee
for a great Workshop full of excellent presentations. I. Bombaci
(U. of Pisa) is also acknowledged for extensive discussions and
insightful ideas on dense matter that lead to improve this talk
substantially. I also thank the warm hospitality and scientific
advise of Dr. G. Lugones on these topics.

% BibTeX users please use
%\bibliographystyle{spmpsci}


\clearpage

\begin{figure}
\plotone{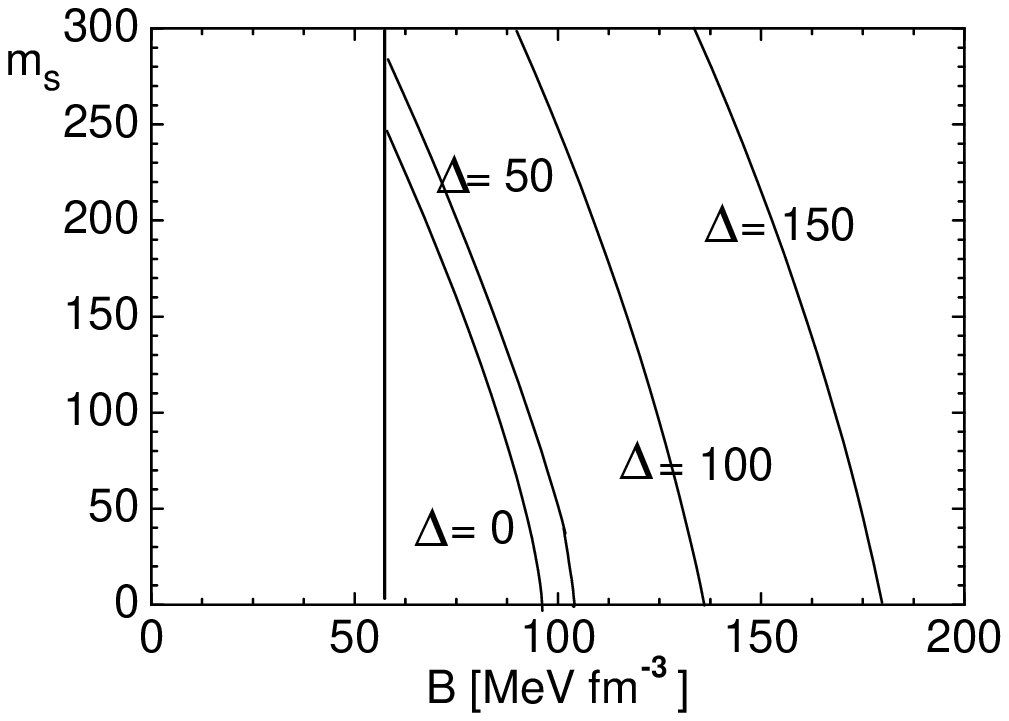} \caption{The stability windows for CFL
strange matter. If the strange quark mass $m_{s}$ and the bag
constant $B$ lie inside the bounded region the CFL state is
absolutely stable. Each window is plotted for $P = 0$, the
stability window is the region between the vertical line (obtained
by requiring instability of two-flavor quark matter) and the curve
with a given value of the gap $\Delta$ as indicated by the label.
Note the enlargement of the window with increasing $\Delta$. See
Lugones \& Horvath (2002) for details.}
\end{figure}

\clearpage

\begin{figure}

\plotone{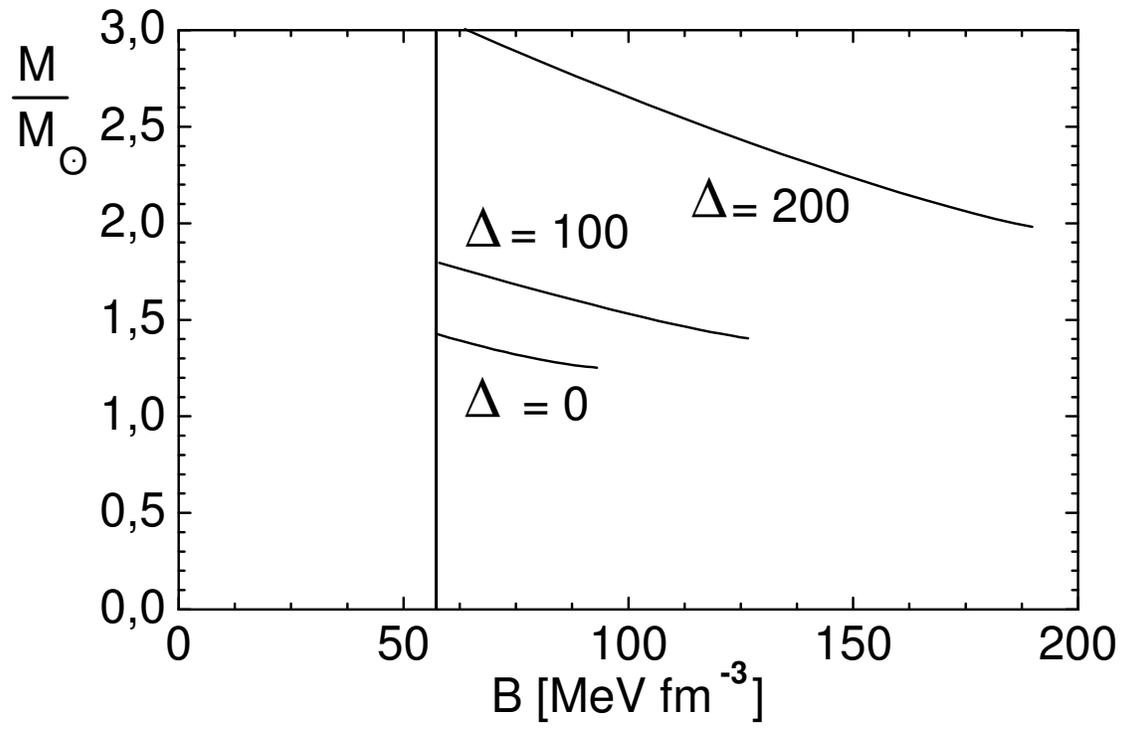}\caption{The maximum mass of each stellar
sequence is shown here as a function of $B$, for $m_s = 150$ MeV
and different values of $\Delta$, the range of the latter is the
same as in Fig.1.}
\end{figure}

\end{document}